\documentclass{aastex631}

\shorttitle{ATA 3I/ATLAS}
\shortauthors{Sheikh et al.}

\usepackage[nolist]{acronym}
\usepackage{xcolor}
\usepackage{longtable}
\usepackage{amsmath}
\usepackage{amssymb}

\definecolor{sof}{rgb}{0.5, 0.0, 0.5}

\begin{document}

\turnoffeditone

\newcommand{\seti}{SETI Institute, 339 Bernardo Ave, Suite 200, Mountain View, CA 94043, USA}

\newcommand{\BLBerkeley}{Breakthrough Listen, University of California, Berkeley, 501 Campbell Hall \#3411, Berkeley, CA 94720, USA}

\newcommand{\BLOxford}{Breakthrough Listen, University of Oxford, Denys Wilkinson Building, Keble Road, Oxford OX1 3RH, UK}

\title{A Search for Radio Technosignatures from Interstellar Object 3I/ATLAS \\ with the Allen Telescope Array}

\author[0000-0001-7057-4999]{Sofia Z. Sheikh}
\affiliation{\seti}
\affiliation{Berkeley SETI Research Center, 339 Campbell, Berkeley, CA 94720}

\author[0009-0000-2879-6539]{Valeria Garcia Lopez}
\affiliation{Furman University, Physics Department, 3300 Poinsett Highway, Greenville, SC 29613.}
\affiliation{\BLBerkeley}

\author[0009-0003-9610-5068]{Isabel Gerrard}
\affiliation{\BLOxford}
\affiliation{\seti}

\author[0000-0002-0637-835X]{James R. A. Davenport}
\affiliation{Department of Astronomy and the DiRAC Institute, University of Washington, 3910 15th Avenue NE, Seattle, WA 98195, USA}

\author[0000-0002-0161-7243]{Wael Farah}
\affiliation{\seti}

\author[0009-0002-2532-3079]{Blayne Griffin}
\affiliation{\seti}
\affiliation{University of Central Arkansas, Department of Physics, Astronomy, and Engineering, 201 Donaghey Ave, Conway, AR 72035}

\author[0000-0003-4823-129X]{Steve Croft}
\affiliation{\BLOxford}
\affiliation{\seti}
\affiliation{\BLBerkeley}

\author[0000-0001-5576-2254]{Luigi F. Cruz}
\affiliation{\seti}

\author[0000-0002-4278-3168]{Imke de Pater}
\affiliation{Department of Astronomy, University of California, Berkeley, 501 Campbell \#3411, Berkeley, CA 94720}
\affiliation{Department of Earth and Space Science, University of California, Berkeley, 501 Campbell \#3411, Berkeley, CA 94720}

\author[0009-0009-6231-9280]{Ben Jacobson-Bell}
\affiliation{Berkeley SETI Research Center, 339 Campbell, Berkeley, CA 94720}
\affiliation{Department of Astronomy, University of California, Berkeley, 501 Campbell \#3411, Berkeley, CA 94720}

\author{Mark Masters}
\affiliation{\seti}

\author[0000-0002-6341-4548]{Karen I. Perez}
\affiliation{\seti}

\author[0000-0002-3430-7671]{Alexander W. Pollak}
\affiliation{\seti}

\author{Carol Shumaker}
\affiliation{\seti}

\author[0000-0003-2828-7720]{Andrew Siemion}
\affiliation{\seti}
\affiliation{Sub-Department of Astrophysics, University of Oxford, Banbury Road, OX1 3RH}
\affiliation{Berkeley SETI Research Center, 339 Campbell, Berkeley, CA 94720}
\affiliation{Jodrell Bank Centre for Astrophysics (JBCA), Department of Physics \& Astronomy, Alan Turing Building, The University of Manchester, M13 9PL, UK}
\affiliation{Department of Physics, Faculty of Science, University of Malta, Msida MSD 2080, Malta}

\begin{abstract}

In 2025 July, the third-ever interstellar object, 3I/ATLAS, was discovered on its ingress into the Solar System. Similar to the NASA Voyager missions sent in 1977, science probes by extraterrestrial life (``artifact technosignatures'') could be sent to explore other stellar systems like our own. In this campaign, we used the SETI Institute's Allen Telescope Array to observe 3I/ATLAS from 1--9\,GHz. We detected nearly 74 million narrowband hits in 7.25\,hr of data using the newly-developed search pipeline \texttt{bliss}. We then \edit1{blanked hits by frequency and drift rate} to mitigate \ac{RFI} in our dataset, narrowing the dataset down to $\sim$2 million hits. These hits were further filtered by the localization code \texttt{NBeamAnalysis}, and the remaining 211 hits were visually inspected in the time-frequency domain. We did not find any signals worthy of additional follow-up. Accounting for the Doppler drift correction and given the non-detection, we are able to set an \ac{EIRP} upper limit of $10-110$\,W on radio technosignatures from 3I/ATLAS across the frequency and drift rate ranges covered by our survey.

\end{abstract}

\acresetall

\section{Introduction} 
\label{sec:intro}

\subsection{3I/ATLAS}
\label{ssec:3I_ATLAS}

On 2025 July 1, the third-ever interstellar object---originally reported as ``C/2025 N1 (ATLAS)'' and now known more commonly as ``3I/ATLAS''---was discovered by the Asteroid Terrestrial-impact Last Alert System (ATLAS) facility in Rio Hurtado, Chile, with the pre-discovery observations extending as far back as 55 days \citep{feinstein2025}. As of 2025 October, the object's orbital eccentricity was estimated to be $e$=6.137\footnote{Ephemerides in \url{https://ssd.jpl.nasa.gov/horizons/}}, making it more eccentric, \edit1{and having a higher interstellar inbound velocity} than the first two known interstellar objects, 1I/`Oumuamua and 2I/Borisov \citep[e.g.,][]{bannister2019natural}. Many telescopes across the world have tracked the orbit of 3I/ATLAS, and ongoing monitoring has revealed clear cometary activity \citep{jewitt2025NOT, alarcon2025TTT, minev2025NAO, belyakov2025palomar, bolin2025apache,chandler2025rubin,rahatgaonkar2025}. If this object is a comet, as expected from the initial characterization, it should be volatile-rich and may produce a dramatic tail now that it has passed perihelion.

In early observations, the object appeared red in color \citep{alvarez2025x, opitom2025musevlt, seligman2025atlas}, with a coma of cometary emission developing as it approached the Sun \citep{cordiner2025,rahatgaonkar2025}. It reached perihelion ($q \sim 1.356$\,AU) on UT 2025 October 29 \citep{seligman2025discovery}, though \edit1{observations from Earth were highly impacted by the small angle to the Sun from late September until early November}. There is still much to learn about this interstellar object, so it is scientifically advantageous that we can observe it throughout its approach. A tentative rotation period of 16.79\,hr has been determined from early observations \citep{de-la-fuente-marcos2025}, which can be compared to 1I/`Oumuamua's very clear tumbling period of 7.3\,hr \citep{fraser2018tumbling} and 2I/Borisov's more ambiguous rotation constraint \citep{bolin2020constraints}. \edit1{OH emission lines have been detected from the object with MeerKAT \citep{pisano2025bmeerkat}, but no continuum emission has yet been reported at radio wavelengths, nor can we constrain the continuum emission from our data.} Early observations are critical for precisely constraining the orbit and any non-gravitational accelerations, as well as the composition of cometary ejecta from 3I/ATLAS.

\subsection{Connection to Technosignatures}
\label{ssec:artifact_technosignatures}

A ``technosignature'' is an astronomically-observable sign of non-human technology built \edit1{by} \ac{ETI}. Historically, most searches for technosignatures have searched for directed electromagnetic radiation, particularly at radio wavelengths \citep[][]{Drake1961}. These radio technosignatures could be efficient for communicating information across interstellar distances, easily distinguishable from natural sources, and relatively unaffected by gas and dust along the line of sight \citep{Cocconi1959}. Recent radio technosignature searches have been covering the immense parameter space more rapidly than ever before \citep[\edit1{e.g.,}][]{choza2023breakthrough}, but the prevalence of radio transmitters in the Galaxy and the universe is still poorly constrained at this point in time \citep{wright2018much}.

Proposals to search for artifacts in the Solar System appeared in the technosignature literature almost simultaneously with radio searches \citep{Bracewell1960}. This is because, despite their association with science fiction, physical artifacts such as probes, surface structures, or spacecraft within the Solar System are as worthy of rigorous characterization and observation as electromagnetic technosignatures at interstellar distances. \citet{sheikh2020nine} employed the Nine Axes framework to evaluate the comparative advantages and disadvantages of artifacts in our Solar System (``Solar System technosignatures,'' in that work), summarized as follows. 

Most fundamentally, Voyager and similar probes will eventually become interstellar objects in other stellar systems. We thus know that no extrapolation is needed for the idea of interstellar technological objects, as we have a proof-by-existence. Passive Solar System artifacts also would not require constant power influx to keep them detectable as technosignatures, unlike electromagnetic transmitters, as they could be unique in observable features (shape/color/composition). This makes them low-cost for the signaler, and likely to be longer-lived as a detectable signature. At merely interplanetary distances (as opposed to interstellar), astronomical facilities are sensitive to smaller objects and less powerful transmissions, making our understanding more complete, and Solar System technosignatures, as a class, more detectable. If the artifact were meant to be discovered as a beacon or a message, it could be packed with data storage in physical form, which has benefits in volume and redundancy over transmitted electromagnetic waves \citep{gertz2016probes}. Finally, Solar System artifacts may also confer unique advantages beyond their use for communication \citep[e.g., \edit1{\acp{ETI}} could leverage stars as gravitational lenses by placing small communication relay artifacts across many stellar systems;][]{kerby2021stellar}.

To differentiate a natural \ac{ISO} from an artificial one, we would need significant evidence of artificiality. Technosignature search strategies for \ac{ISO}s over the next decade are outlined in detail by \citet{davenport2025}. Observables could include non-gravitational trajectories beyond those expected from natural cometary emission \citep[e.g.][]{taylor2024}, unusual bolometric optical properties (e.g., reflectance, albedo), spectral lines inconsistent with a standard cometary or asteroidal composition, unusual morphologies or rotation profiles, or other anomalous features \citep{rogers2024weird}. The most unambiguous observable, however, would be narrowband emission, particularly in the radio, which can only be produced by technological sources. While some technological artifacts may have no radio emission at all (e.g., Voyager in the far future), many of the potential motivations for creating and sending objects across interstellar distances would benefit strongly from an artificial radio component. 

We can thus use high frequency-resolution radio observations to characterize potential emission from \ac{ISO}s. We seek to understand the population-level characteristics of \ac{ISO}s, i.e., being able to set limits on the percentage of these objects that host radio transmitters. Even if no technosignatures are detected, high frequency resolution radio spectroscopy can also be leveraged to detect cometary emission \citep{crovisier2016,park2018,pisano2025ameerkat} to better understand the properties of natural \ac{ISO}s. 

As established above, the technosignature hypothesis is not unreasonable in a galaxy with other technologically-capable life (which we may or may not reside in). We note that there is no compelling evidence yet that 3I/ATLAS has any non-natural characteristics \edit1{despite} extensive observation by the scientific community. However, it is still worth setting constraints on technological radio emission from 3I/ATLAS throughout its passage of our Solar System. This is standard practice \citep[e.g., see][for a search for narrowband radio emission from 1I/`Oumuamua]{enriquez2018breakthrough}, and should be performed for all \ac{ISO}s in the future as well.

\subsection{The Allen Telescope Array}
\label{ssec:ata}

The \ac{ATA} is a 42-element radio interferometer at \ac{HCRO} in Hat Creek, California. The 6.1-m offset Gregorian elements are in the process of being refurbished with dual-polarization log-pyramidal ``Antonio'' feeds. These feeds are cryocooled to 70\,K, allowing for sensitivity from 1--10\,GHz. The current active array consists of the 28 of the 42 dishes that have been upgraded to date. The analog signals from these feeds are amplified and sent over optical fiber to the on-site signal processing room. There, the signals are heterodyned with four independently-tuned local oscillators, with each of these independent tunings sampling 672\,MHz anywhere in the available frequency range of the instrument. Finally, data from the four tunings are digitized and saved to the network via a 100-Gb ethernet link\footnote{This has since been upgraded to 400\,Gb, after the observations reported here}. Further details on the \edit1{digital signal processing} chain and hardware refurbishment process will be described by \citet{ata-dsp} and  \citet{ata-instrumentation}.

In Section~\ref{sec:observations}, we will discuss the observing parameters and campaign choices for the 3I/ATLAS \ac{ATA} observations. In Section~\ref{sec:analysis}, we outline the data analysis and hit filtering that were used to search for signals-of-interest. The signals that we visually inspected are discussed in Section~\ref{sec:results} and the implications for the presence of narrowband transmitters are covered in Section~\ref{ssec:upper_limits}. Finally, we summarize and conclude in Section~\ref{sec:conclusions}.

\section{Observations}
\label{sec:observations}

The \ac{ATA} is privately owned and operated by the SETI Institute, allowing for rapid follow-up of targets-of-opportunity between larger \ac{SETI} surveys. Following the announcement of 3I/ATLAS, the ATA team rapidly designed an observing campaign and initiated on-sky observations on 2025 July 2, approximately 23 hours following the initial ATLAS report. In general, the frequency of a potential \ac{ETI} transmitter is unknown, and covering more parameter space is more valuable than trying to guess at ``magic frequencies'' \citep{zuckerman1985magic}. Given this, we leveraged the \ac{ATA}'s large bandwidth and multiple simultaneous tunings to maximize our frequency coverage. The sessions were divided into three frequency chunks: ``low'' (1000--3688\,MHz), ``mid'' (3688--6376\,MHz) and ``high'' (6376--9064\,MHz). Each chunk consisted of four \edit1{consecutive} 672-MHz tunings. We took data across five observing sessions, with a total of 7.25\,hr of on-source observations. Details for each observing session are shown in Table~\ref{tab:observing_log}.

\begin{table}[ht!]
    \centering
    \begin{tabular}{|c|c|c|}
    \hline
        Date of First Scan  &   Frequency Ranges & Total 
        Observing Time \\
        (UTC) & (MHz) & (hr) \\
    \hline
        2025-07-03-05:10:38 &   1000--3688    &  1.3\\
        2025-07-05-04:50:14 &   3688--6376   &  1.1\\
        2025-07-08-04:52:40 &   6376--9064    &  3.4\\
        2025-07-09-05:20:16 &   1000--3688    &  1.1\\
        2025-07-10-05:16:28 &   3688--6376   &  0.3\\
    \hline
    \end{tabular}
    \caption{Observing sessions for this campaign, comprising 7.25 hours over 1--9\,GHz.}
    \label{tab:observing_log}
\end{table}

During the sessions, a series of 5-minute beamformed scans of 3I/ATLAS were taken using the in-house beamformer backend \texttt{BLADE} \citep{ata-blade}. \texttt{BLADE} is flexible for different science cases, and includes options for both high frequency resolution and high time resolution modes. In this project, we used the high frequency resolution mode of \texttt{BLADE} which allows the user to specify the exact frequency and time resolution, the polarization data to be saved (if any), and the number and location of beamformed beams. The data are saved as filterbank HDF5 files across 96-MHz sub-bands, with seven sub-bands per tuning (or 28 \edit1{sub-bands} per frequency ``chunk'' as described above). For this project, we chose a frequency resolution of 1.9\,Hz (sufficiently fine to resolve narrowband signals) and a sampling time of 16.8\,s. We saved only the Stokes-I product (i.e., no polarization information was recorded). Two beams were placed on the sky for each scan: one ``on-beam'' at the center of the incoherent beam, centered on the location of 3I/ATLAS, and another ``off-beam'' five synthesized beamwidths away, with the inter-beam distance calculated for the center frequency of each tuning and applied to the entire tuning.

Prior to each observing session, the \ac{ATA}'s GPU-based cross-correlation backend was used to target 3C\,286, a flux and phase calibrator which is stable across the entire frequency range of the \ac{ATA} \citep{perleybutler2017flux}. These calibration scans lasted 10 minutes each. We then used our in-house \texttt{CASA} \citep{mcmullin2007casa} pipeline to calculate the instrumental delays, phases, and bandpass solutions which were then applied to the subsequent beamformer observations. The data taken for this campaign totaled $\sim$22\,TB. 

\section{Data Analysis}
\label{sec:analysis}

\subsection{Finding narrowband hits with \texttt{bliss}}
\label{ssec:bliss}

We used the code \texttt{bliss}\footnote{\edit1{The software is available on GitHub \texttt{bliss} codebase: \url{https://github.com/UCBerkeleySETI/bliss} under a 3-Clause BSD License. The version used for this analysis is archived in Zenodo \citep{nathan_west_2025_bliss}.}} to search for Doppler-drifting narrowband signals within the 7.25\,hr of data from the campaign described in Section~\ref{sec:observations}. \texttt{bliss} is a successor to the \texttt{turboSETI} \citep{enriquez2017breakthrough} and \texttt{seticore}\footnote{\edit1{The software is available on GitHub \texttt{seticore} codebase: \url{https://github.com/lacker/seticore} under an MIT License.}} software packages. \edit1{\texttt{bliss} is the newest and most feature-rich implementation, and addresses many of the issues with \texttt{turboSETI} raised by \citet{choza2023breakthrough} and \citet{margot21}. From the 2D dynamic spectrum (time--frequency) plane, \texttt{bliss} calculates the drift--frequency plane by de-drifting the dynamic spectrum to a range of drift rate values and summing over time, then searches for signals by finding local maxima in the drift--frequency plane. This is more computationally intensive than the fast method used by \texttt{turboSETI}, which is based on a method for pulsar dedispersion by \citet{taylor74}, but by calculating the full drift--frequency plane, \texttt{bliss} can find overlapping signals with different drift rates. \texttt{bliss} also takes into account signal widths for resolved signals, which enables it to more accurately estimate the power of signals at large drift rates compared to the unit drift rate or ``one-to-one point'' \citep{sheikh2019choosing}. A pre-search step removes the effect of the coarse channelization, further improving the code's signal power estimation.} For a more thorough description of the \texttt{bliss} code, see \citet{sheikh2025ataasp} and \citet{jacobson-bell2025bliss}. 

We ran the \texttt{bliss\_find\_hits} utility from \texttt{bliss} with a minimum drift rate of $-4$\,Hz/s, a maximum drift rate of 4\,Hz/s, and a \ac{SNR} threshold of 15. This drift rate range is relatively standard in the field and is sufficient to identify signals at a plausible range of radial accelerations relative to the receiver (see further drift rate discussion for this particular object in Section~\ref{ssec:blanking}). Similarly, an \ac{SNR} threshold of 15 \edit1{balances sensitivity with mitigation of false positives from \ac{RFI} at \ac{HCRO}. That said, it should be noted that this survey uses one of the lowest \ac{SNR} thresholds in any modern \ac{SETI} survey. Some previous \texttt{turboSETI} searches, including \citet{price20breakthrough} and \citet{choza2023breakthrough}, have used a nominal \ac{SNR} threshold of 10, but as \citet{choza2023breakthrough} show, all \texttt{turboSETI} searches to date require an \ac{SNR} threshold correction factor of 3.3, implying that the true \ac{SNR} threshold of \citet{price20breakthrough} and \citet{choza2023breakthrough} is 33. The SNR estimation of \texttt{bliss} is much more accurate \citep{jacobson-bell2025bliss}, so our nominal \ac{SNR} threshold of 15 is approximately the true \ac{SNR} threshold. Consequently, it is somewhat expected that this search will return a high number of hits due to its increased algorithmic sensitivity to fainter signals.}

\subsection{RFI blanking and drift rate limits}
\label{ssec:blanking}

\begin{figure}[ht!]
    \centering
    \includegraphics[width=\textwidth]{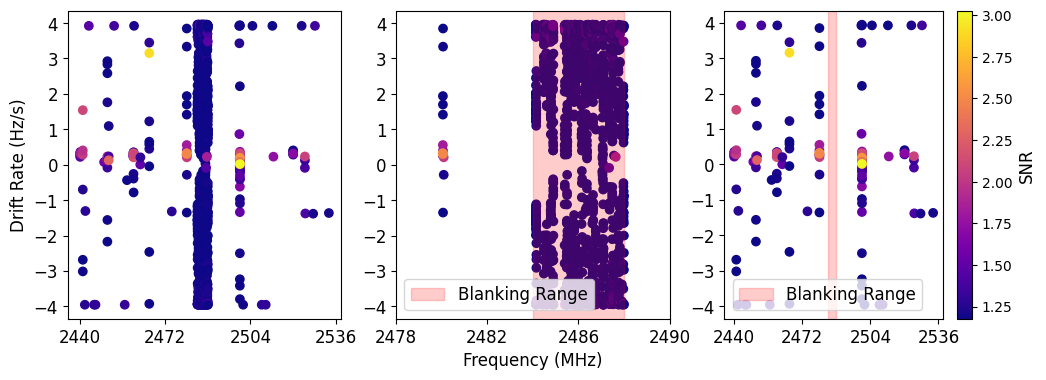}
    \caption{\edit1{
        An example of using \texttt{bliss} results to identify \ac{RFI} blanking ranges. All three sub-panels depict one sub-band (2440--2536\,MHz), shown in the frequency (x-axis, in MHz) and drift rate (y-axis, in Hz/s) plane. [Left] All hits detected by \texttt{bliss}, before any filtering, in a 5-minute scan (1712 hits). [Middle] A zoomed in view of 2484--2488\,MHz, a region visually identified as a high density region across drift rates and therefore flagged as \ac{RFI} (e.g. 1565 hits in the shaded range). [Right] Identical to the left figure, but with the blanking range from the middle figure applied to the data, leaving only 147 hits after filtering.}}
    \label{fig:flag_blanking}
\end{figure}

By covering 1--9\,GHz, we take advantage of the \ac{ATA}'s wide bandwidth, especially in the less-explored high frequency regions (i.e., above 5\,GHz) which have not been as well-covered in prior \ac{SETI} searches. However, this wide bandwidth also includes the spectrally-contaminated L-band, which covers downlink frequencies for GPS and other persistent classes of \ac{RFI}; this comes with its own challenges, especially in an era of increasing human radio emissions. Trying to run the signal localization code \texttt{NBeamAnalysis}\footnote{\edit1{The software is available on GitHub \texttt{NBeamAnalysis} codebase: \url{https://github.com/isabelgerrard/NBeamAnalysis} under a 3-Clause BSD License and version 2.0 is archived in Zenodo} \citep{ntusay_igerrard_nbeam}.} without first filtering the data would be prohibitively expensive in terms of computation time. Thus, to minimize the effect of these high-interference spectral windows on our data analysis, we ``blank'' any spectral region with a large amount of hits, removing all hits in those boundaries from further analysis\footnote{We hope to apply more sophisticated \ac{RFI} removal in these regions in the future, \edit1{for example, clustering with more signal parameters (e.g., signal width, \ac{SNR}) to specifically target known transmitters}} but for now the number of hits returned is computationally prohibitive. \edit1{Frequencies were flagged for blanking by visually inspecting the \texttt{bliss} outputs of a random subset of scans from each 96 MHz sub-band and identifying intervals with an apparent high density of hits across all drift rates searched ($\pm 4$\,Hz/s) that is persistent for at least 0.5\,MHz (Figure \ref{fig:flag_blanking}).} A saturation of hits spanning a wide range of drift rates for an extended bandwidth suggests that an underlying signal was not a strong match to a single drift rate and thus is a strong indicator of local interference. The full list of blanked frequency intervals is shown in the appendix (Table~\ref{tab:blanking_ranges}) and in sum constitutes 16\% of the total 1--9\,GHz band.

\begin{figure}[ht!]
    \centering
    \includegraphics[width=\textwidth]{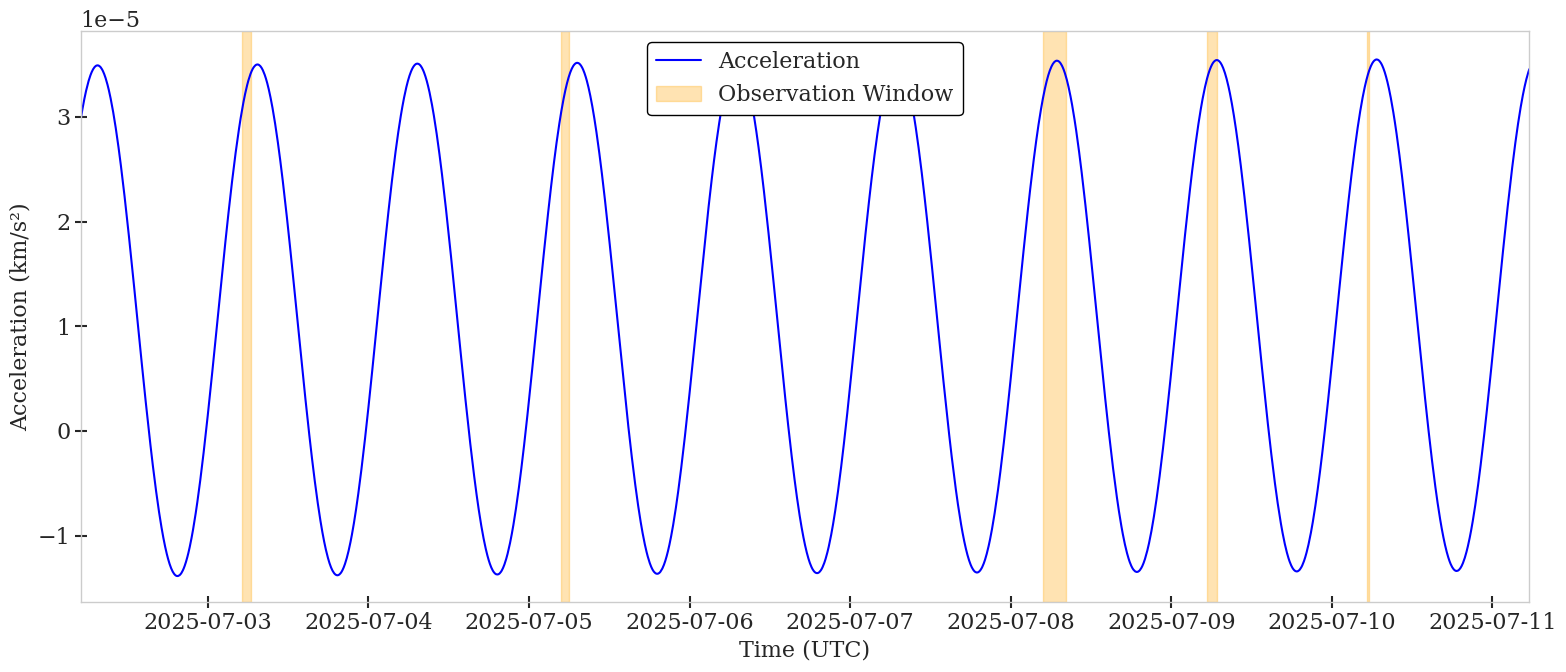}
    \label{fig:acceleration}
    \caption{The radial acceleration of 3I/ATLAS over the $9$ days containing our observations (blue line). Observation times are highlighted in yellow. This allows us to determine the expected drift rates for potential signals coming from 3I/ATLAS.}
\end{figure}

To filter out signals that do not match the dynamics of 3I/ATLAS, we characterized the acceleration of the object during the times of observation, to compare the observed drift rates with those expected from the acceleration between the object and the Earth. We took the time derivative of the \texttt{deldot} parameter within JPL Horizons, which gives the apparent range rate relative to the observer in km/s. Figure \ref{fig:acceleration} displays the resulting acceleration as a function of time, which takes a sinusoidal form due to the radial acceleration contribution from the Earth's rotation. From this, we estimated the average and standard deviation of the acceleration during the observation intervals. The values for all five sessions fell within $3.2 \times 10^{-5}$\,km/s$^2$ to $3.5 \times 10^{-5}~\textrm{km/s}^2$, with very low standard deviation within the sessions ($10^{-7}$\,km/s$^2$), indicating that the same filter can be used for all sessions. This is logically consistent, given that the sessions began slightly after the object's rise time each night, implying a similar geometrical contribution from Earth's rotation each night.

We then converted the acceleration to a drift rate using the following equation from \citet{sheikh2019choosing}:

\begin{equation}
    \dot{\nu} = \frac{\frac{dv}{dt} \times \nu_{obs}}{c}
    \label{eq:drift_rate}
\end{equation}

where $\dot{\nu}$ is the drift rate in Hz/s, $\nu_{obs}$ is the frequency in Hz, $\frac{dv}{dt}$ is the acceleration as calculated previously, and $c$ is the speed of light. Drift rates in Hz/s are dependent on observing frequency; to set a threshold for the entire search range in Hz/s, we need to consider the Hz/s drift rate of the highest frequencies in the survey. At a center frequency of 8728\,MHz (our highest tuning), the maximum drift rate of 3I/ATLAS would be 1.01\,Hz/s. Meanwhile, at a center frequency of 1336\,MHz (our lowest tuning), the minimum drift rate of 3I/ATLAS would be 0.15\,Hz/s. We considered applying drift-rate-specific filters for each of the tunings to incorporate this information in a frequency-dependent way. However, the rotation or tumbling period of 3I/ATLAS is not yet certain: this contribution, especially in Solar System objects, can be extremely large. \citet{sheikh2019choosing} note that asteroid 2008\,DP4 tumbles in a way that could result in observed drift rates up to 4.22\,nHz (i.e., 4.22\,Hz/s at 1\,GHz). Here, we take a middle ground approach. We filter our hits for drift rates between 0.05--1.5\,Hz/s at all frequencies, sufficient to capture the known acceleration of 3I/ATLAS at all frequencies within our sample with some flexibility for additional rotational contributions\footnote{One might expect an active artifact to have stabilization against rotation, much like our telescopes and spacecraft. However, passive artifacts may have no such stabilization. To complicate matters further, one could imagine that rotation could be used as a tool to create artificial gravity within a space environment \citep{polstorff1965dynamics}}. 

\subsection{Search for localized events with \texttt{NBeamAnalysis}}
\label{ssec:NBeam}

After running \texttt{bliss} and filtering out hits based on the frequency and drift rate cuts described in Sections~\ref{ssec:bliss} and \ref{ssec:blanking}, the remaining hits were run through the \texttt{NBeamAnalysis} spatial filtering pipeline. This pipeline is described in detail by \citet{tusay2024radio}, but we will describe it briefly here. In Section~\ref{sec:observations}, we described the two-beam observing strategy that we employed using the ATA beamformer mode: the on-beam was placed on the target (3I/ATLAS) and the off-beam was placed 5 synthesized beamwidths away. The \texttt{NBeamAnalysis} pipeline employs a series of different procedures which leverage two beams to identify signals that are consistent with a localized point source on the sky:
\begin{itemize}
    \item Compare the hit tables from \texttt{bliss} for the on-beam and the off-beam. If two hits in the tables are nearly identical in frequency and have an \ac{SNR} ratio between the two beams that is less than $\sqrt{N_{antennas}}$ = $\sqrt{28}$ = 5.29, then the hits are removed. \edit1{Note that the threshold of $\sqrt{N_{antennas}}$ is conservative: it is the theoretical ratio between the power of a signal in an incoherent beam versus in a coherent beam, not a comparison of two coherent beams (which, in this case, would be expected to be $>>$5.29 except in the sidelobes, but dependent on the frequency- and pointing-dependent beam pattern). This method has been tested with true positive signals (orbiter downlinks from Mars) by \citet{tusay2024radio} where the $\sqrt{N_{antennas}}$ threshold was deemed sufficient to recover known signals as expected without inconvenient levels of false positives.}
    \item \edit1{Calculate the dot product or ``DOT score'' between the two data arrays surrounding the hits.} If the DOT score is 1, the beams are identical. DOT scores close to 0 indicate two arrays with different properties, potentially identifying signals that are present much more strongly in the on-beam.
    \item Calculate the SNR ratio between the two arrays around every hit. As in the first step, any hit with an SNR ratio greater than 5.29 is potentially of interest.
\end{itemize}

Up to 1000 plots depicting signals with SNR ratios greater than 5.29 are generated for visual inspection. If more than 1000 signals surpass that threshold, then the 1000 with the lowest DOT scores are plotted. If no signals have SNR ratios greater than 5.29, the 100 signals which score best in a combined assessment of the DOT score and the SNR ratio are plotted for visual inspection.

\section{Results}
\label{sec:results}

\begin{figure}[ht!]
    \centering
    \includegraphics[width=0.6\textwidth]{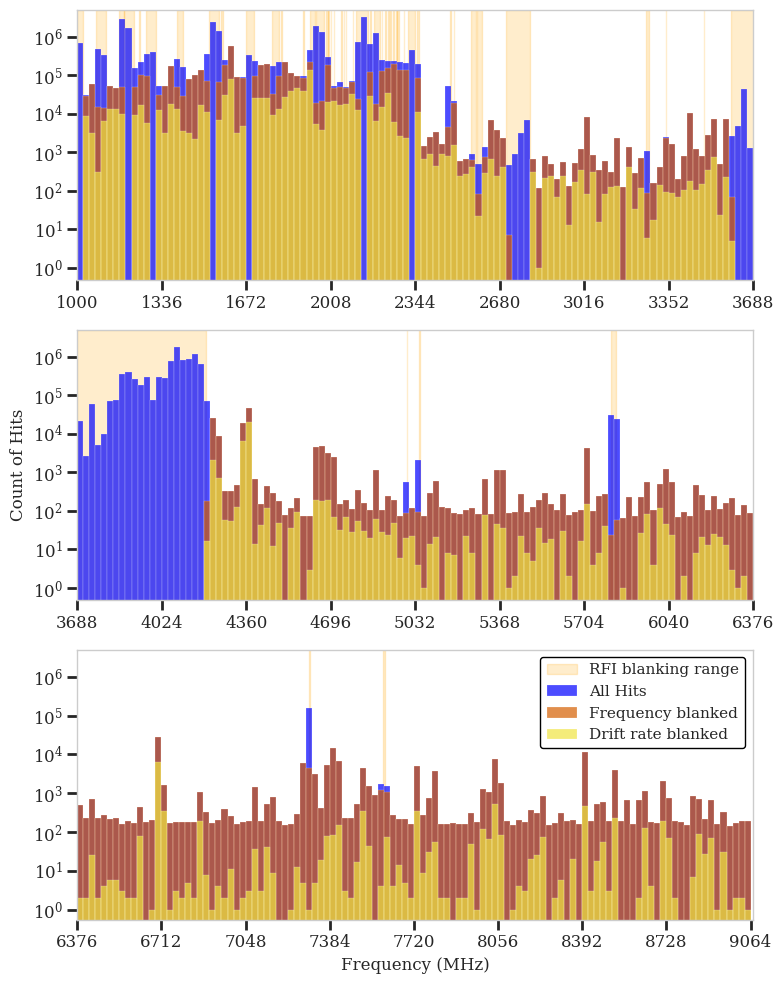}
    \caption{Frequency distribution of the $\sim74$ million hits obtained in this survey. The top, middle, and bottom panels correspond to the ``low'' (1000--3688\,MHz), ``mid'' (3688--6376\,MHz), and ``high'' (6376--9064\,MHz) frequency ranges, respectively. All y-axes are scaled to the same limits. All hits identified by \texttt{bliss} are shown in blue, blanking ranges are shown with \edit1{yellow shading behind the histogram}, the hits after blanking within the frequency ranges are shown in brown, and the hits after limiting the drift rate range are shown in gold. Applying blanking ranges and drift rate limits significantly reduced dense clusters of hits (reducing the total number of hits by 97.4\%) which improved our ability to process the data and detect putative signals associated with 3I/ATLAS.
}
    \label{fig:freq_histogram}
\end{figure}

\begin{figure}[ht!]
    \centering
    \includegraphics[width=\textwidth]{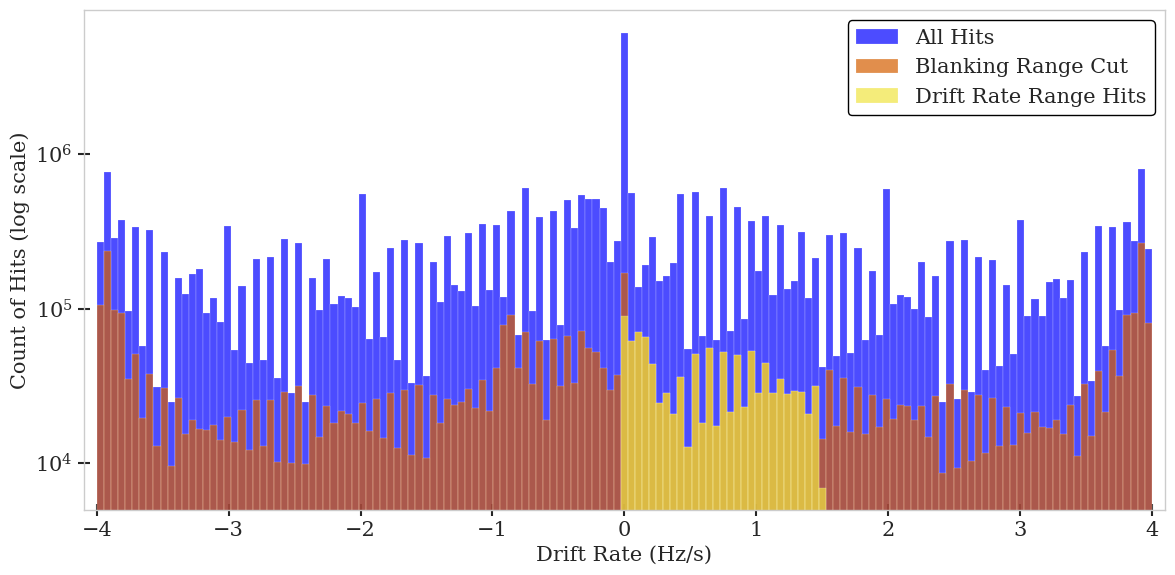}
    \caption{The distribution of the $\sim 74$ million hits from this survey across drift rate. Note the peak of hits around 0\,Hz/s (expected because most \ac{RFI} transmitters are in the same reference frame as the telescope, i.e., on the ground). After the drift rate cut, the count of hits is much more even across the restricted drift rate range. Note that the sawtooth pattern seen across drift rate is a known artifact in the current development version of \texttt{bliss}, caused by an overestimation of \ac{SNR} for high-bandwidth ($\gtrsim 10$\,Hz wide) hits close to integer multiples of the unit drift rate. It causes increased amounts of false positives close to these drift rates but does not cause false negatives \citep[for more details, see][]{jacobson-bell2025bliss}.}
    \label{fig:all_drift_rate}
\end{figure}

\begin{figure}[ht!]
    \centering
    \includegraphics[width=\textwidth]{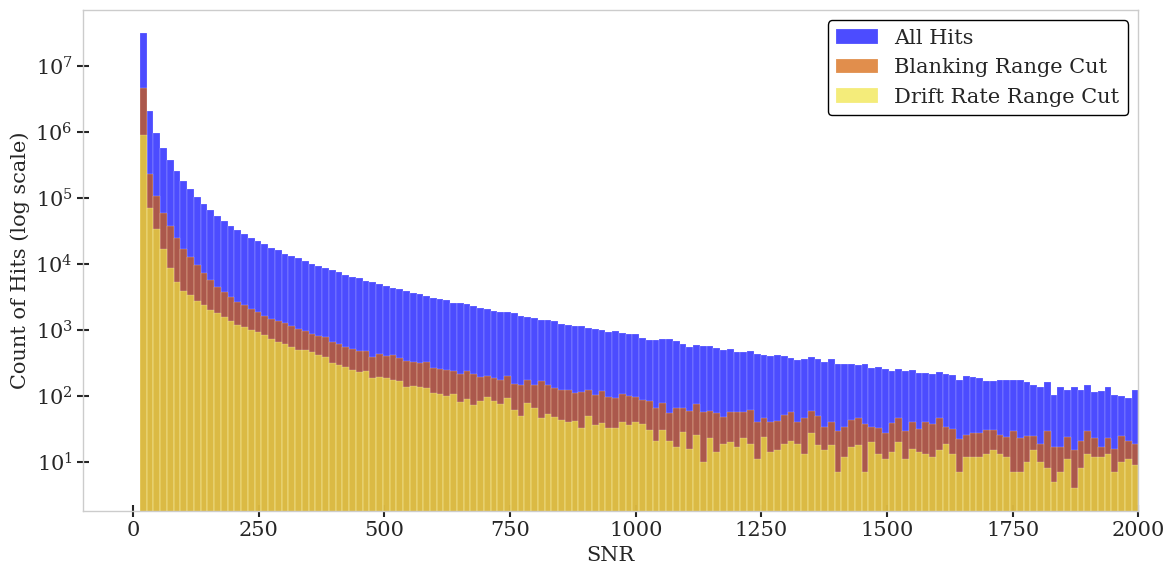}
    \caption{The distribution of the $\sim 74$ million hits from this survey across \ac{SNR}. As expected, there are significantly more weak hits than strong ones, and there is no significant trend in \ac{SNR} when filters are applied to frequency and drift rate.}
    \label{fig:SNR}
\end{figure}

After running \texttt{bliss} on all of our observations, we identified a total of 73,974,799 narrowband hits across the 7.25\,hr of observation. After removing any hit in the frequency blanking ranges from Table~\ref{tab:blanking_ranges}, which sum to a total blanked bandwidth of 1269.8\,MHz ($\sim$16\% of the total bandwidth) we were then left with $\sim$11~million hits. After drift rate blanking, $\sim$2 million hits remained. The frequency distribution of this population of hits, before and after the blanking steps, \edit1{is shown in Figure~\ref{fig:freq_histogram}. The same population is plotted in drift rate in Figure~\ref{fig:all_drift_rate} and in \ac{SNR} in Figure~\ref{fig:SNR}.} \edit1{Although this process was done without actively incorporating domain knowledge of \ac{FCC} allocations or spectrum management, the regions of the spectrum with significant numbers of hits (more than a million in a single bin in Figure~\ref{fig:freq_histogram}) can be attributed to known interferers such as GPS L5 (blanking range of 1168--1184.3\,MHz), the INMARSAT downlink (blanking range of 1525--1562\,MHz), LTE Bands 1 and 2 downlinks (2110--2154.6\,MHz and 1951--1980\,MHz blanking ranges, respectively), and the C-band geostationary satellite downlink (blanking range of 3700--4200\,MHz). This confirms that the method of identifying regions of interference in \texttt{bliss} outputs from Figure~\ref{fig:flag_blanking} worked as expected.}

\begin{figure}[ht!]
    \centering
    \includegraphics[width=\textwidth]{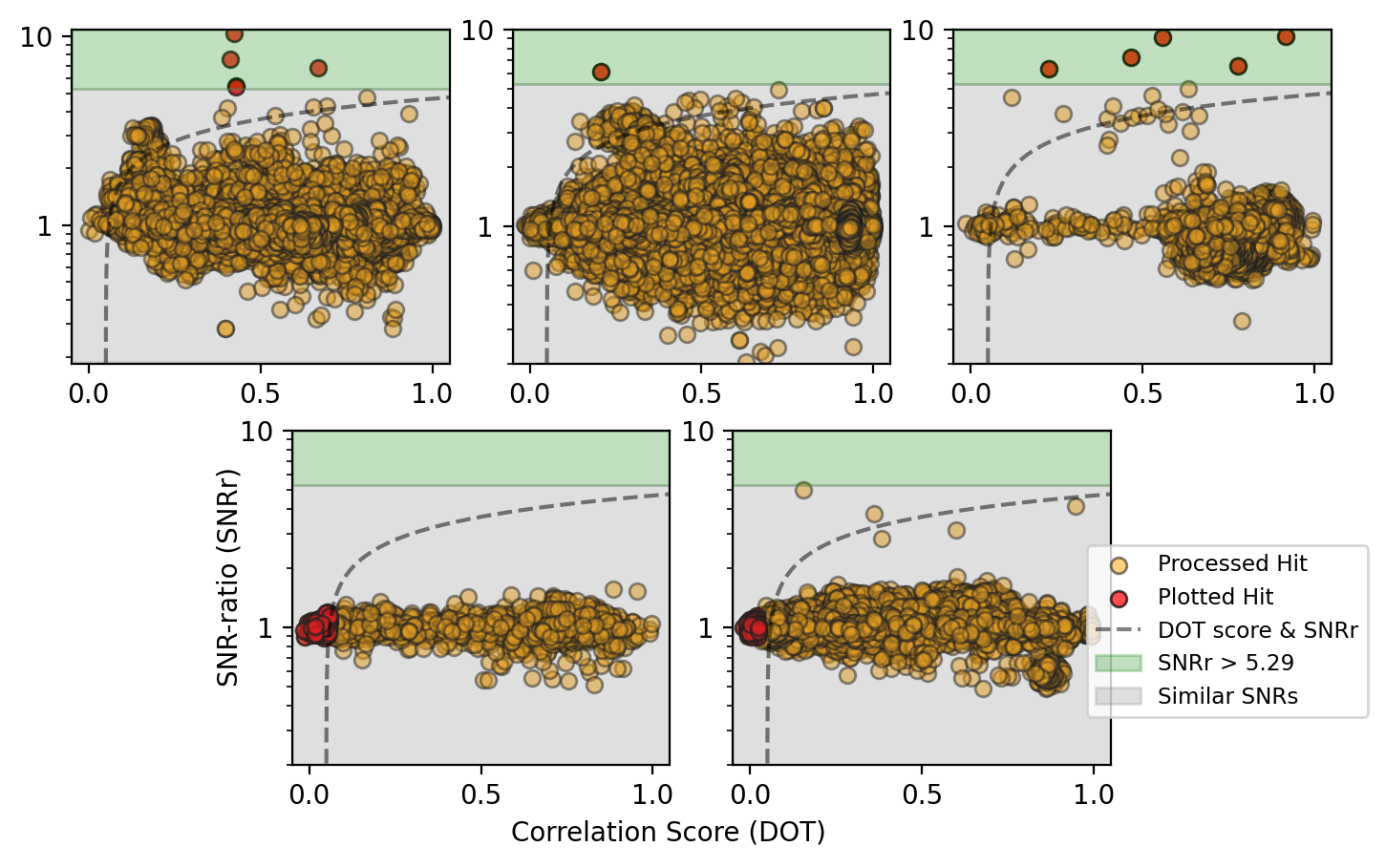}
    \caption{\edit1{
        DOT score versus \ac{SNR} ratio plots for each of the five observing sessions (one per sub-plot), that depict the population of hits processed by \texttt{NBeamAnalysis} and the subset that were plotted for visual inspection. For sessions that had hits with \ac{SNR} ratios that exceeded the threshold of 5.29 (top three sub-plots), only the 11 total points in the green regions were plotted for visual inspection. For the two sessions that had no signals that exceeded this ratio (bottom two sub-plots), the 100 hits with the best DOT score that exceeded the empirical function given by \citet{tusay2024radio} were selected.}}
    \label{fig:snr_v_dot}
\end{figure}

\begin{figure}
    \centering
    \includegraphics[width=\linewidth]{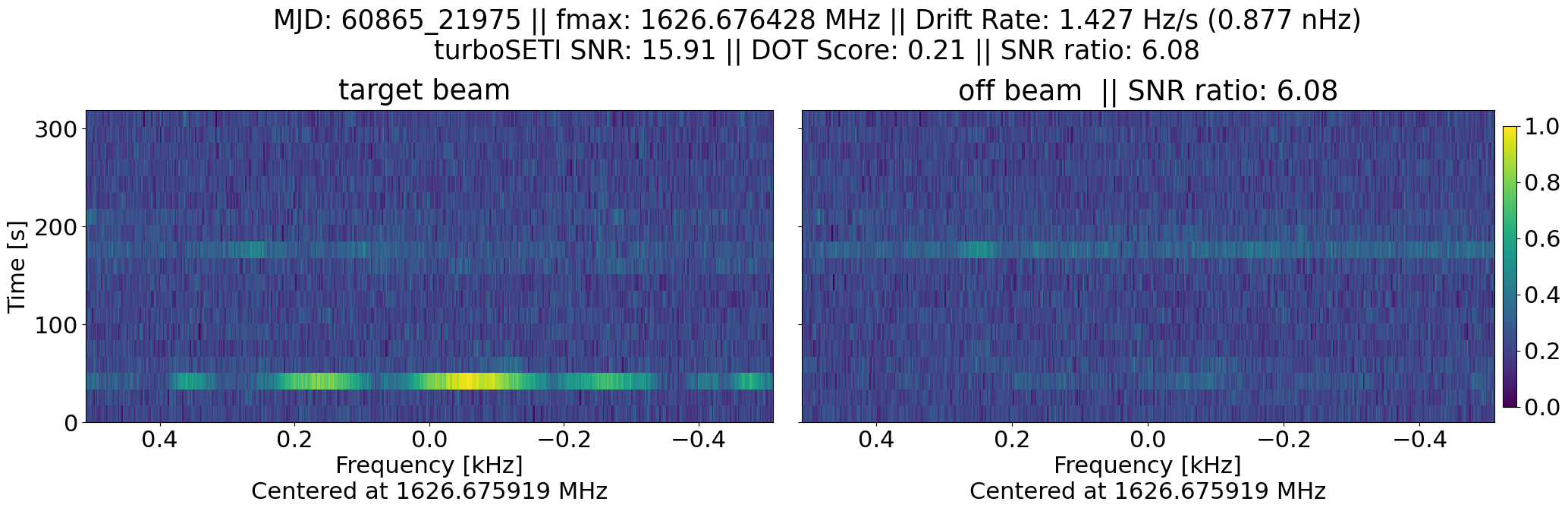}
    \caption{An output waterfall plot from \texttt{NBeamAnalysis} for a signal which was ranked in the top 211 events. The on-beam (pointed at 3I/ATLAS) is shown in the left subplot, while the off-beam is shown in the right subplot. Each subplot is a normalized waterfall plot with frequency on the horizontal axis, time on the vertical axis, and intensity (scaled to the on-beam plot from 0--1) indicated by the color bar. This hit had a frequency outside of the blanking ranges, a drift rate in the allowable range, a relatively low DOT score (indicating significant difference between the plots), and an SNR ratio above 5.29 (indicating consistency with a point source on the sky). However, two things immediately discount this signal as an \ac{ETI} technosignature: (1) the signal in the on-beam is not a narrowband technosignature despite the \texttt{bliss} hit, but rather some kind of broader-band modulated signal, and (2) this frequency is consistent with the L-band downlink for the Iridium satellites. The Iridium \ac{RFI} explanation is consistent with the point source characteristics of the hit as well, as satellites and other distant human transmitters can mimic the characteristics that we would expect from a true \ac{ETI} technosignature.}
    \label{fig:example_event}
\end{figure}

\begin{figure}
    \centering
    \includegraphics[width=\linewidth]{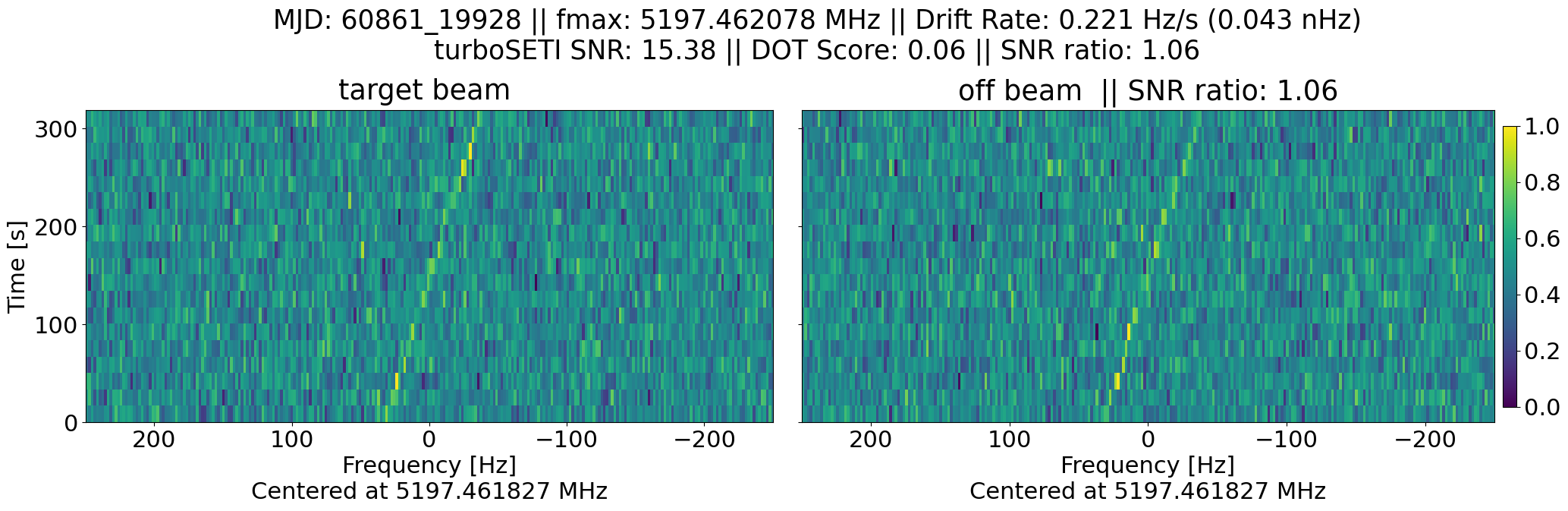}
    \caption{An output waterfall plot from \texttt{NBeamAnalysis} from an observation with 0 signals above the 5.29 SNR ratio threshold. This signal had a non-blanked frequency, a drift rate in the allowable range, and does appear to be truly narrowband (i.e., technological). However, the signal has a similar \ac{SNR} in both the on-beam and the off-beam, indicating a local interferer in the allocation allotted to fixed-satellite service (Earth-to-space).}
    \label{fig:example_event_2}
\end{figure}

After running \texttt{NBeamAnalysis}, 211 plots were highlighted for further analysis. \edit1{All of the events that were plotted are consistent with standard scatter in the \ac{SNR}-ratio vs. DOT score space---the plane that defines the anomaly score of the remaining events and prioritizes events for plotting (see Figure~\ref{fig:snr_v_dot}).} Of these plots, 11 are events with \ac{SNR} ratios greater than 5.29 \edit1{signifying the event was brighter when pointed at the target location (potentially sky-localized), and likely indicating the signal source was not on the surface of the Earth (``non-terrestrial'').} The remaining 200 plots were from two observations that did not produce any final signals above the threshold, resulting in the backup plotting functionality, which selects 100 signals with the best combined assessment of DOT score and \ac{SNR} ratio. After visual inspection, all of these 211 events were easily attributable to RFI: \edit1{none of the 11 sky-localized events depict narrowband drifting signals, and the narrowband drifting signals in the remaining 200 plots are not consistent with a sky-localized signal coming from 3I/Atlas, therefore zero outputs depict non-terrestrial technological signals.} Two examples of those 211 events are shown in Figures~\ref{fig:example_event} and \ref{fig:example_event_2}.

A summary of the number of hits that passed each stage of analysis is shown in Table~\ref{tab:hit_filtering}.

\begin{table}[ht!]
    \centering
    \begin{tabular}{|c|c|c|c|c|c|c|}
    \hline
    \textbf{Date (UTC)} & \textbf{\texttt{bliss}} & \textbf{RFI Blanking} & \textbf{Drift Rate Limits} & \textbf{\texttt{NBeamAnalysis}} & \textbf{Plotting} & \edit1{Signals-of-Interest}\\
    \hline
    2025-07-03 & 32,346,482 & 5,887,236 & 823,046 & 288,316 & 5 & 0\\ 
    2025-07-05 & 12,415,496 & 63,451 & 4,339 & 1,330 & 100 & 0\\ 
    2025-07-08 & 609,692 & 626,114 & 22,675 & 8,332 & 100 & 0\\ 
    2025-07-09 & 23,511,437 & 4,363,855 & 1,020,006 & 448,305 & 1 & 0\\ 
    2025-07-10 & 5,091,692 & 234,930 & 63,716 & 29,557 & 5 & 0\\ 
    \hline
    \textbf{Total} & 73,974,799 & 11,175,586 & 1,933,782 & 775,840 & 211 & 0\\  
    \hline
    \end{tabular}
    \caption{The number of hits remaining after each filtering step of the pipeline. The stark differences between different dates are due both to the variable observing session length and the different amount of \ac{RFI} in each frequency band. Note that the two nights that returned 100 plots were returning the default number of ``backup'' plots with no plots surpassing the SNR ratio threshold i.e., they were generally of less interest.}    
    \label{tab:hit_filtering}
\end{table}

\section{Upper Limits}
\label{ssec:upper_limits}

Given 3I/ATLAS's distance from Earth during the observing window ($\sim$3.35\,AU), the ATA would be able to detect a narrowband radio signal if its flux density exceeds a threshold $S_{min}$ as follows:

\begin{equation}
    S_{\text{min}} = \frac{S/N_{\text{min}}}{\sqrt{\beta}} \cdot \left( \frac{S_{\text{sys}}}{\Delta \nu_t} \right) \cdot \sqrt{\frac{\Delta \nu}{n_p \cdot \tau_{\text{obs}}}}
\end{equation}

Here, $S/N_{\text{min}}$ is the desired signal-to-noise threshold, $S_{\text{sys}}$ is the system equivalent flux density (in e.g., Jy), $\Delta \nu_t$ is the transmitter bandwidth, $\Delta \nu$ is the channel bandwidth of the data, $n_p$ is the number of polarizations, $\tau_{\mathrm{obs}}$ is the observing time, and $\beta$ is the dechirping efficiency. This equation is slightly modified from the \texttt{turboSETI} formulation in e.g., \citet{gajjar2021Breakthrough} in the following way. Given our frequency resolution ($\Delta \nu \approx 1.9~\text{Hz}$) and temporal resolution ($\Delta t \approx 16.777~\text{sec}$), signals with high frequency drift are spread across multiple channels, resulting in reduced sensitivity: this will happen to some degree regardless of the de-drifting algorithm being applied. However, compared to the \texttt{turboSETI} de-Doppler algorithm considered by \citet{gajjar2021Breakthrough}, the \texttt{bliss} algorithm recovers $\sqrt{\beta}$ by including neighboring bins containing smeared signal into the power estimate.

This dechirping efficiency can be expressed as follows for our survey:

\begin{equation*}
\beta = [\mathrm{round}\big(\frac{\Delta t}{\Delta\nu} \times |\dot{\nu}|\big)]^{-1}
\end{equation*}

The parameter $\dot{\nu}$ represents the drift rate being searched. Currently \texttt{bliss} rounds $|\dot{\nu}|$ to the nearest integer multiple of the unit drift rate $\frac{\Delta\nu}{\Delta t}$, accounting for the difference between this formulation of $\beta$ and the one in \citet{gajjar2021Breakthrough}. Our system achieves maximum dechirping efficiency for drift rates of 0.113 Hz/s or less, with sensitivity getting gradually worse as the drift rates increase. Based on our minimum $S_{\mathrm{min}}$ at various frequencies, we can then calculate the corresponding \ac{EIRP} for all our observations using:

\begin{equation}
    \text{EIRP}_\text{obs} = S_\text{min} \times 4\pi D^2_\text{obs} \times \Delta \nu
\end{equation}

We also calculated a worst-case \ac{EIRP} for the lowest dechirping efficiencies in our survey (i.e., those corresponding to 1.5\,Hz/s), modifying the flux sensitivity accordingly. The \ac{EIRP} values across the frequencies from 1 to 9\,GHz are shown in Table~\ref{tab:sefd_values}.

The beamformer \ac{SEFD} values shown in Table~\ref{tab:sefd_values} are derived from the \ac{SEFD}s of individual antennas, calculated by CASA during calibration (as described in Section~\ref{sec:observations}). We remove outliers by identifying antennas in each polarization with \ac{SEFD}s greater than 1.5 times the interquartile range above the third quartile; antennas with abnormally high \ac{SEFD} values generally have hardware issues at the feed level and should be removed from the dataset (in this survey, this affected 8--12 of the 56 antenna-polarization combinations depending on the frequency). Then we take an unweighted mean of individual antenna \ac{SEFD}s and divide by the number of elements remaining in the list for that polarization. Finally, the two average polarization \ac{SEFD} values are themselves averaged to obtain the final beamformer \ac{SEFD}, and, for frequencies with two sessions, those two session \ac{SEFD}s are averaged as well to give a single beamformer \ac{SEFD} for each frequency tuning. Note that the error bars on the beamformer \ac{SEFD} values in Table~\ref{tab:sefd_values} are derived from the standard deviation of the population of individual antenna SEFDs, which fall around 15\%: this is relatively standard for the ATA and these uncertainties can be propagated linearly to the $S_{\min}$ and \ac{EIRP} values if desired. 

\begin{table}[ht!]
    \centering
    \begin{tabular}{|c|c|c|c|}
    \hline
    \textbf{Center Frequency} & \textbf{SEFD} & \textbf{$S_{\min}$} & \textbf{EIRP} \\
    (MHz) & (Jy) & (Jy) & (W) \\
    \hline
    1336 & 390 $\pm$ 60 & 170--2300 & 10--40 \\ 
    2008 & 420 $\pm$ 56 & 190--2400 & 11--40 \\ 
    2680 & 460 $\pm$ 54 & 200--2700 & 12--40 \\ 
    3352 & 460 $\pm$ 60 & 200--2700 & 12--40 \\ 
    4024 & 560 $\pm$ 68 & 250--3200 & 15--50 \\ 
    4696 & 620 $\pm$ 84 & 280--3600 & 16--60 \\ 
    5368 & 640 $\pm$ 114 & 280--3700 & 17--60 \\ 
    6040 & 670 $\pm$ 118 & 300--3900 & 18--60 \\ 
    6712 & 740 $\pm$ 105 & 330--4300 & 20--70 \\ 
    7384 & 810 $\pm$ 139 & 360--4700 & 21--80 \\ 
    8056 & 940 $\pm$ 225 & 420--5500 & 25--90 \\ 
    8728 & 1150 $\pm$ 326 & 500--6700 & 31--110 \\ 
    \hline
    \end{tabular}
    \caption{The \ac{SEFD} of the beamformer, minimum detectable flux density ($S_{\min}$), and \ac{EIRP} values for different center frequencies. The ranges in $S_{\min}$ and \ac{EIRP} demonstrate values for the best-case to worst-case dechirping efficiencies.}    
    \label{tab:sefd_values}
\end{table}

Finally, we can compare these \ac{EIRP} values with the recently-posted values from \citet{pisano2025ameerkat}: in the narrowband technosignature search of 3I/ATLAS in that work, the authors report a minimum \ac{EIRP} value of 0.17\,W. This value is about commensurate with a cellphone handset, significantly more sensitive than this work; this is expected given MeerKAT's world-class sensitivity. However, MeerKAT only set limits from 900--1670\,MHz, so this work increases the frequencies covered for 3I/ATLAS by a factor of 10.

\acresetall

\section{Conclusions} 
\label{sec:conclusions}

In this work, we considered the most recently-discovered interstellar object, 3I/ATLAS, through a technosignature lens, motivating this frame in Section~\ref{sec:intro}. We took 7.25\,hr of data on 3I/ATLAS across 1--9\,GHz with the \ac{ATA}, using a high frequency resolution beamforming mode designed for narrowband signal searches (see Section~\ref{sec:observations}). In Section~\ref{sec:analysis}, we described the deployment of the narrowband hit-finding code \texttt{bliss}, \ac{RFI} blanking and drift rate limits, and spatial filtering with the \texttt{NBeamAnalysis} code to filter our initial $\sim 74$ million \texttt{bliss} hits down to 211 plots for visual inspection. We found no evidence of narrowband technosignatures in this dataset (Section~\ref{sec:results}). This result implies \ac{EIRP} upper limits of $10-110$\,W across our frequency and drift rate range. 

In the future, the blanking ranges for \ac{RFI} in this project will be explored in more detail to determine whether modulation-sensitive analysis strategies could mitigate the \ac{RFI} while still allowing for narrowband searches in these ranges. In addition, as far as we know, this is the first time that expected drift rate contributions have been used as a filter for a technosignature search of a Solar System object. These advances in technosignature searches with \ac{ATA} will be applied to future searches on the instrument, further pushing into unexplored parameter space for the \ac{SETI}.

\appendix

\begin{table}[ht!]
    \centering
    \label{tab:blanking_ranges}
    \begin{tabular}{|l|l|l|l|l|}
    \hline
    1000--1025   & 1775--1802   & 2062.3--2062.8 & 2257--2258      & 2705--2800   \\
    1075--1116   & 1811--1813   & 2067.3--2067.8 & 2269--2270.3    & 3264--3274   \\
    1168--1184.3 & 1899--1901   & 2097--2098     & 2271.3--2279    & 3341--3343   \\
    1187--1227   & 1927--1945   & 2110--2154.6   & 2299--2301      & 3492--3493   \\
    1246--1250   & 1951--1980   & 2166--2169.5   & 2316--2345      & 3600--4200   \\
    1275--1314   & 1990--1996   & 2179--2203.5   & 2353--2359.5    & 4999--5000.9 \\
    1395--1420.5 & 1999--2000.5 & 2211--2212     & 2484--2488      & 5048--5049.5 \\
    1525--1562   & 2010--2011   & 2226--2227.5   & 2497.25--2498.5 & 5810--5830   \\
    1572--1580   & 2022--2023   & 2237--2241     & 2565.75--2585   & 7300--7307   \\
    1671--1705   & 2049--2054   & 2244--2248     & 2590--2609      & 7596.5--7604 \\
    \hline
    \end{tabular}
    \caption{The blanking ranges, in MHz, used for this survey. Any hit from \texttt{bliss} that had a center frequency within a blanking range was removed from the survey due to high levels of \ac{RFI} at those frequencies.}
\end{table}

The Allen Telescope Array (ATA) refurbishment program and its ongoing operations have received substantial support from Franklin Antonio. Additional contributions from Frank Levinson, Greg Papadopoulos, the Breakthrough Listen Initiative and other private donors have been instrumental in the renewal of the ATA. The Paul G. Allen Family Foundation provided major support for the design and construction of the ATA, alongside contributions from Nathan Myhrvold, Xilinx Corporation, Sun Microsystems, and other private donors. The ATA has also been supported by contributions from the US Naval Observatory and the US National Science Foundation. Breakthrough Listen is managed by the Breakthrough Initiatives, sponsored by the Breakthrough Prize Foundation. BG acknowledges support from the SETI Institute REU program (NSF Award 2447895). JRAD acknowledges support from the DiRAC Institute in the Department of Astronomy at the University of Washington. The DiRAC Institute is supported through generous gifts from the Charles and Lisa Simonyi Fund for Arts and Sciences, Janet and Lloyd Frink, and the Washington Research Foundation.

\begin{acronym}
    \acro{SETI}{Search for Extraterrestrial Intelligence}
    \acro{ETI}{Extraterrestrial Intelligence}
    \acro{ATA}{Allen Telescope Array}
    \acro{RFI}{Radio Frequency Interference}
    \acro{SNR}{Signal-to-Noise Ratio}
    \acro{FRB}{Fast Radio Burst}
    \acro{BLADE}{Breakthrough Listen Accelerated Digital signal processing  Engine}
    \acro{FWHM}{full width at half maximum}
    \acro{BL}{the Breakthrough Listen Initiative}
    \acro{HCRO}{Hat Creek Radio Observatory}
    \acro{GBT}{Green Bank Telescope}
    \acro{FCC}{Federal Communications Commission}
    \acro{SEFD}{System Equivalent Flux Density}
    \acro{EIRP}{Effective Isotropic Radiated Power}
    \acro{ISO}{Interstellar Object}
\end{acronym}

\bibliography{bibliography}
\bibliographystyle{aasjournal}

\end{document}